\documentstyle[aas2pp4,psfig]{article}

\slugcomment{Revised 19 July 1999}

\lefthead{Smale}
\righthead{LMC~X--2 a $Z$-source?}

\def\source{LMC~X--2}
\def\approxlt{\mathrel{\hbox{\rlap{\lower.55ex \hbox {$\sim$}}
        \kern-.3em \raise.4ex \hbox{$<$}}}}
\def\approxgt{\mathrel{\hbox{\rlap{\lower.55ex \hbox {$\sim$}}
        \kern-.3em \raise.4ex \hbox{$>$}}}}

\begin{document}

\title{LMC~X--2: The First Extragalactic $Z$-Source?}

\author{Alan P. Smale\altaffilmark{1}}
\affil{Laboratory for High Energy Astrophysics,
Code 662, NASA/Goddard Space Flight Center, Greenbelt, MD 20771}

\author{Erik Kuulkers}
\affil{Space Research Organization Netherlands, Sorbonnelaan 2, 3584
CA Utrecht\altaffilmark{2}}

\altaffiltext{1}{Also Universities Space Research Association.} 

\altaffiltext{2}{Also Astronomical Institute, Utrecht University,
P.O. Box 80000, 3507 TA Utrecht, The Netherlands.} 

\begin{abstract}

We present RXTE observations of \source\ obtained during a five-day
interval in 1997 December, during which the source was radiating at a
mean intensity near the Eddington limit, and strongly variable on timescales
of seconds to hours. The shapes of the X-ray color-color and
hardness-intensity diagrams during the observations, the presence of
VLFN and HFN in the power spectra, and the high intrinsic X-ray
luminosity of \source\ (which historically spans 0.4--2.0$L_{Edd}$ for
reasonable estimates of the neutron star mass) are more characteristic
of a $Z$-source in its flaring branch than of an atoll-source.  On
this basis, we provisionally reclassify \source\ as a $Z$-source, the
eighth such source known and the first to be detected beyond our
Galaxy.

Using periodogram and Fourier analysis of the X-ray lightcurve we
detect an apparently-significant modulation with a period of
8.160$\pm$0.011 hrs and a semi-amplitude increasing from 14\% in the
1.8--4.0 keV range to 40\% at 8.7--19.7 keV.  This X-ray modulation
appears to confirm a candidate orbital periodicity determined from
optical photometry ten years prior to our campaign, but we cannot
rule out a chance alignment of intrinsic X-ray flares. Current RXTE
ASM light curves and archival EXOSAT observations show no sign
of such a pronounced periodicity.

The X-ray spectrum of \source\ can be well fit using variations of
simple Comptonization models. Fits to
phase($\simeq$intensity)-resolved spectra show strong correlations
between the power law slope (in one parameterization) or the depth to
optical scattering (in another) and phase. We discuss the implications
of these results for the inclination, geometry, and emitting regions
of the \source\ system.

\end{abstract}


\section{Introduction}

Considering its extreme brightness, it is perhaps surprising how
little is known about \source\ (0521$-$720). One of the most luminous
persistent sources in the Large Magellanic Cloud, it is the only one
of these original five bright X-ray sources to be identified as a low
mass X-ray binary (LMXB).  With an observed X-ray luminosity in the
range 0.6--3$\times$10$^{38}$ erg~s$^{-1}$ (van Paradijs 1995),
\source\ has a persistent emission which can exceed the Eddington
luminosity for a neutron star with a canonical mass of 1.4--2.0M$_{\sun}$,
making it one of the most powerful persistently-emitting low mass
X-ray binaries. Because of its accurately-known distance and low
line-of-sight absorption, studies of \source\ can provide not only a
clear view of the physics and dynamics of such a system, but also an
important comparison with Galactic LMXBs.

The spectrum of the $V$=18.8 optical counterpart (Johnston et al.\ 1979)
shows the expected strong He~II emission overlying a blue continuum,
but the N~III--C~III--O~II blend emission typically seen at $\lambda
\lambda$4640-4650\AA\ is weak or non-existent (Bonnet-Bidaud et al.\ 1989;
Crampton et al.\ 1990). The Bowen blend is also absent in the optical
spectra from the supersoft LMC sources, indicating that the lack of
this emission is linked to the relatively low metallicity of the
LMC. This underabundance may also contribute to the high luminosities
observed in the LMC X-ray sources in general, compared to the Galactic
population. Accretion of material onto compact objects may be
limited by preheating due to photoelectric absorption of X-rays by the
heavier elements in the material, thus sources in a lower-abundance
environment might be expected to accrete closer to the Eddington
limit.

For \source\ the ratio of X-ray to optical luminosities is
$L_x/L_{opt}\sim$600, similar to that seen in Galactic LMXBs. The
weakness of the 6.7 keV Fe line (White et al.\ 1986; Bonnet-Bidaud et al.\
1989) supports the lower-abundance argument.  Searches for the orbital
period of the system have produced contradictory results, with
evidence for photometric and/or spectroscopic periods of 6.4 hr
(Bonnet-Bidaud et al.\ 1989), 8.16 hr (Callanan et al.\ 1990), and 12.5 days
(Crampton et al.\ 1990). Thus, even the nature of the companion and the
scale of the binary system are unclear. A 6--8 hr period would imply
a late-type dwarf companion, while the longer 12.5-day period would
require an evolved companion in a much larger orbit (e.g.\ Hasinger \& van
der Klis 1989).

Searches of the literature and the archival holdings at the HEASARC
reveal that few dedicated observations of \source\ have been
performed. Data from the earlier satellites consist almost entirely of
surveys of (or including) the LMC. Flux measurements prior to 1977,
and the long-term behavior of the source, are summarized in Griffiths
\& Seward (1977). The HEAO-1 and Einstein survey data can be found in
Johnston, Bradt, \& Doxsey (1979) and Long, Helfand, \& Grabelsky
(1981).  A short multiwavelength study of \source\ was
conducted by Bonnet-Bidaud et al.\ (1989), using EXOSAT and the ESO
complex at La Silla in 1984. They determined that the X-ray spectra
could be described using either a Comptonization model, or the sum of
two components, i.e.\ Bremsstrahlung plus a blackbody. In the first
representation, the authors note that their ability to determine the
existence or non-existence of a blackbody component was strongly
limited by the available statistics. In the second, the blackbody
component contributed 20--40\% of the total emission, and the
blackbody radius was found to be 6--10 km.  Variability is typically
observed in the light curve on $\sim$5--30 min timescales, and a flare
(brightness enhancement) was also seen. However, this flare only
represents a 25\% increase in the persistent emission at the end of
the observation, and is not well modeled by an increase in the
blackbody flux in the two-component representation.  The EXOSAT
observations presented by Bonnet-Bidaud et al.\ lasted 8.2 hours,
too short to effectively search for a period in the 4--12 hr range or
beyond. No clear correlation was observed between X-ray and optical
emission, and no significant QPO or red noise features were detected
in the 2--64~Hz range studied.

\begin{deluxetable}{ll}
\tablewidth{0pc}
\tablecaption{Observation Log}
\tablehead{
\colhead{Observation} &
\colhead{Start/Stop Time (UT)}
}

\startdata
1 & 1997 Dec 2 18:52 -- 1997 Dec 3 02:00 \nl
2 & 1997 Dec 3 19:07 -- 1997 Dec 4 02:00 \nl
3 & 1997 Dec 4 18:53 -- 1997 Dec 5 01:59 \nl
4 & 1997 Dec 5 19:20 -- 1997 Dec 6 02:00 \nl
5 & 1997 Dec 6 18:51 -- 1997 Dec 7 01:40 \nl
\enddata
\end{deluxetable}

Accreting LMXBs can be divided into two classes based on their
fast-timing behavior and variations in their spectral color or
hardness (Hasinger \& van der Klis 1989; for a review, see e.g.\ van
der Klis 1995). The $Z$-sources are named for the
three-branched shape they trace in a diagram of 'hard' color against
'soft' color. From top to bottom, the three branches of the $Z$ are
known as the Horizontal, Normal, and Flaring branches (HB, NB, FB), and it is
believed that the accretion rate $\dot M$ increases as the source
moves from the HB to the FB.

The $Z$-sources generally show quasi-periodic oscillations in their
powers spectra in the 1--100 Hz range whose characteristics are
closely correlated with their position on the $Z$.  Horizontal Branch
QPOs, or HBOs, are probably explained by a magnetospheric beat
frequency model, while Normal/Flaring branch QPOs (N/FBOs) are related
to oscillations in the accretion flow near/beyond the Eddington limit.
In addition, the power spectra show power-law shaped noise at
frequencies 
$\gtrsim$1 Hz (known as very low frequency noise, or VLFN); band-limited
($\sim$1--10~Hz) noise with cut-off frequencies of $\sim$10~Hz (called
low-frequency noise, LFN) in the Horizontal and upper part of the
Normal branch; and high-frequency noise (HFN) with cut-off frequencies in
excess of 50--100 Hz, which sometimes leads to a
broad, peaked appearance.  The currently-known $Z$-sources are
Sco~X--1, GX~17$+$2, GX~349$+$2 (=Sco~X--2), Cyg~X--2, GX~340$+$0, and
GX~5$-$1 (Hasinger \& van der Klis 1989). When bright, Cir~X--1 also
shows $Z$-source characteristics (Shirey et al.\ 1998; Shirey, Bradt
\& Levine 1999).

The second class, the so-called 'atoll' sources, describe a much
simpler curved 'banana' shape in their color-color diagram, sometimes
supplemented with an 'island' of datapoints during low-luminosity
intervals. The atoll sources show VLFN with a generally shallower
power law index than the $Z$-sources, and a broad low-frequency
band-limited noise component often (confusingly) also termed HFN. 

Both $Z$- and atoll sources also display QPO oscillations at
frequencies of 200--1200 Hz (``kiloHertz QPOs'', see e.g. van der Klis
1998) and at 1--60 Hz (see e.g.\ the review of van der Klis 1995 for
the $Z$-sources, and Wijnands \& van der Klis 1999 and references
therein for a tabular summary of such oscillations in the atoll
sources).

To date the $Z$-sources have tended to show longer orbital periods
($\sim$18~hrs -- 10 days), implying evolved companions, while atoll
sources tend to have orbital periods under 10 hrs. Confirmed
membership of one or other class has direct physical consequences for
the system; the $Z$-source characteristics imply the existence of
magnetospheres with magnetic fields of $B\sim$10$^{9-10}$G, while the
magnetic fields in the atoll sources would be substantially lower (see
Psaltis, Lamb, \& Miller 1995). The difference in the nature of the
companions possibly implies that $Z$- and atoll-sources differ in
their evolutionary history (Hasinger \& van der Klis 1989).

Assignation of \source\ to one or other of these classes can only be
achieved by studying the color-color diagram and the fast time
variability of the source.  No clear color-color correlation was
observed during the EXOSAT observations, but it is apparent from the
hardness diagrams presented by Bonnet-Bidaud et al.\ (1989) that the
low-level variability analysis was affected by the counting statistics.
Despite its absolute brightness, the greater distance of \source\
relative to similar Galactic sources results in a paucity of photons
for high-time-resolution studies.

We have therefore undertaken a program of study of \source\ using
RXTE, to search for the orbital period of the system and determine the
signature of its fast time variability. We have also obtained
simultaneous optical coverage at SAAO with the 1.9m telescope and
high-speed CCD photometer, to search for correlated variability
between the X-ray and optical regimes; these data will be presented in
a later publication (McGowan, Charles, \& Smale 1999).

\section{Observations}

We observed \source\ with RXTE (Bradt, Rothschild, \& Swank, 1993) 
between 1997 December 2 18:52 UT --
December 7 1:40 UT, for an on-source total good time of 100 ksec. This
time was divided into five separate observations, in an observing
strategy designed to achieve the maximum amount of simultaneity with
the optical coverage mentioned above. The X-ray data we present
here were obtained using the PCA instrument with the Standard 2 and
Good Xenon configurations, with time resolutions of 16 sec and
$<1\mu$sec respectively.  The PCA consists of five Xe proportional
counter units (PCUs), with a combined effective area of about 6500 cm$^2$
(Jahoda et al 1996). All 5 PCUs were on throughout the
observations, apart from a 6000-sec interval at the beginning of the
third observation during which PCU~4 was turned off. In our light
curves we have included the data from this short interval normalized
to 5 PCUs. 

We performed our data extraction using the RXTE standard data analysis
software, FTOOLS 4.2.  Background subtraction of the PCA data was
performed utilizing the ``skyvle/skyactiv'' models generated by the
RXTE PCA team. The quality of the background subtraction was checked
in two ways: (i) by comparing 
the source and background spectra and light curves
at high energies (50--100 keV) where \source\ itself no longer
contributes detectable events; and (ii) by using the same models to
background-subtract the data obtained during slews to and from the
source. We conclude that our background subtractions in the 2--20 keV
energy range are accurate to a fraction of a count per second.

\begin{figure}[htb]
\figurenum{1}
\begin{center}
\begin{tabular}{c}
\psfig{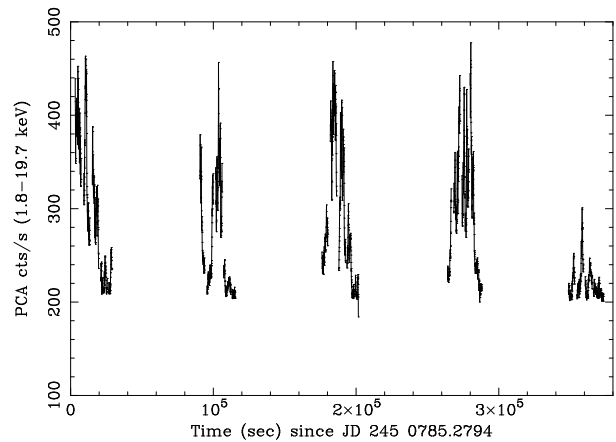}
\end{tabular}
\caption{Background-subtracted X-ray light curve of \source, 1997
December 2--7, extracted using an energy range 1.8--19.7~keV and a time
resolution of 64 sec.}
\end{center}
\end{figure}

\begin{figure}[htb]
\figurenum{2}
\begin{center}
\begin{tabular}{c}
\psfig{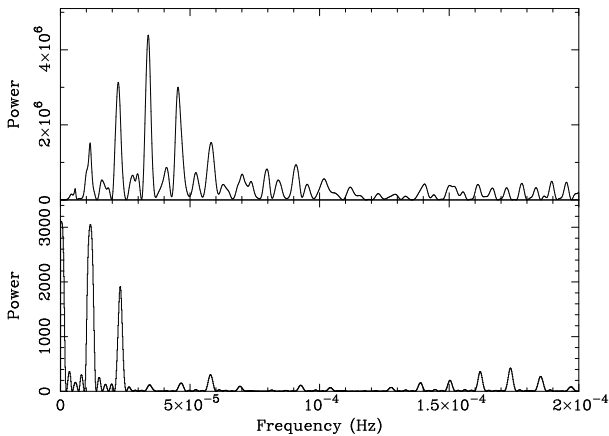}
\end{tabular}
\caption{The upper panel shows the Fourier power spectrum of the
background-subtracted X-ray light curve. The most significant peak
corresponds to a period of 29446$\pm$147s; the peaks on either side
are the one- and two-day aliases of this fundamental period. The
window function, in the lower panel, shows peaks at frequencies
corresponding to periods of 1~days and 0.5~days. Fitting sine waves to the
raw data, we refine the period determination to
$P$=29377$\pm$40s =8.160$\pm$0.011 hrs (see text).}
\end{center}
\end{figure}

We examined the rapid aperiodic variability in \source\ using the high
time resolution PCA data obtained in the Good Xenon mode.  Discrete
power spectral density distributions (PSDs) were calculated by
dividing the data into segments of uniform length, performing fast
Fourier transforms of each, and averaging the results.  The PSDs were
normalized such that their integral gives the squared RMS fractional
variability (Miyamoto et al 1991; van der Klis 1989). We subtracted
the Poisson noise level from the power spectra, taking into account the
modifications expected from PCA detector deadtime. As described by
Zhang et al (1995, 1996) this deadtime comes in two parts: the
$t_d$=10$\mu$s contribution from the events within in the time series
themselves, and a $\tau$=170$\mu$s contribution from energetic cosmic
ray events (very large events, or VLE) (see also Morgan, Remillard, \&
Greiner 1997).

Data were also obtained with the HEXTE phoswich detectors, which are
sensitive over the energy range 15--250~keV (Rothschild et al. 1998).
The source was detected with a background-subtracted count rate of
0.75 count~s$^{-1}$.  However, because of the cut-off spectral shape
of \source, these counts were limited to energies $<$25 keV, and the
HEXTE data did not in this case provide any information additional to
that gained using the PCA.

\begin{figure}[htb]
\figurenum{3}
\begin{center}
\begin{tabular}{c}
\psfig{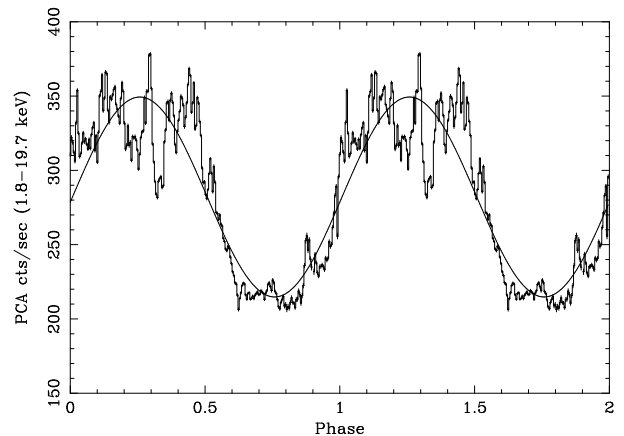}
\end{tabular}
\caption{The X-ray light curve folded on the candidate orbital period
of 29377 sec, with the best-fitting sine wave superimposed over the
data. The fifth data segment was excluded before folding. Two cycles
are repeated for clarity.}
\end{center}
\end{figure}

\section{Results}

\subsection{The light curve and HID/CD}

We show the overall 1.8--19.7~keV X-ray light curve in Fig.~1. Pronounced
variability is apparent, with some evidence for a modulation on an
$\sim$8~hr timescale in the first four observations, and a suggestion
of a similar but much lower-amplitude modulation in the fifth. 
Using a discrete Fourier
transform technique optimized for unevenly-sampled time series
(Scargle 1989), we found that the strongest peaks in the power
spectrum appear at $P$=29446$\pm$147s and the one-day aliases of $P$
(Fig.~2). No other significant periodicities were detected.
The strength and significance of this primary peak is
corroborated with a standard period-folding code utilizing the
L-statistic (Davies 1990), with a false alarm probability of less than
0.01\%. Aware of the pitfalls of detecting orbital periods in
datatrains with individual lengths (in this case averaging 20000s)
similar to the period of interest, we have scrutinized the 
window function of the observations (e.g. van der Klis 1989); from our
Fig.~2 we see that the only principal peaks in the window function
correspond to the one-day and half-day artefacts from the data sampling. 
In addition, we have
created and analyzed a series of 1000 randomized datasets with the
same temporal spacing. None of the randomized datasets showed significant
Fourier power or L-statistic peaks in the range of candidate periods of
interest ($\sim$hrs).

\begin{figure}[htb]
\figurenum{4a}
\begin{center}
\begin{tabular}{c}
\psfig{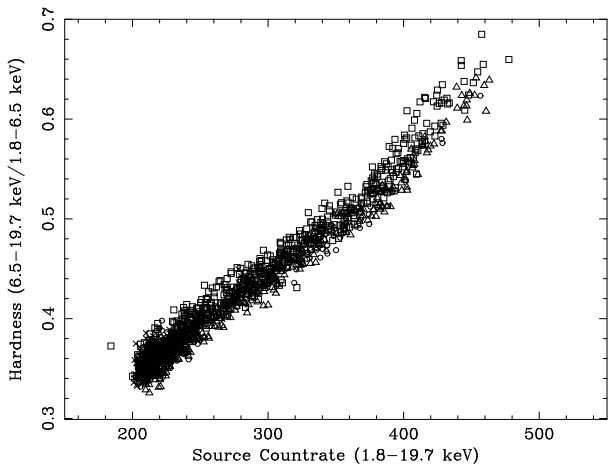}
\end{tabular}
\caption{The hardness/intensity diagram for the observation. 
Data points from the five
observing segments are marked using triangles, circles, squares,
crosses, and filled squares respectively. Each point represents 64s of
data.}
\end{center}
\end{figure}

\begin{figure}[htb]
\figurenum{4b}
\begin{center}
\begin{tabular}{c}
\psfig{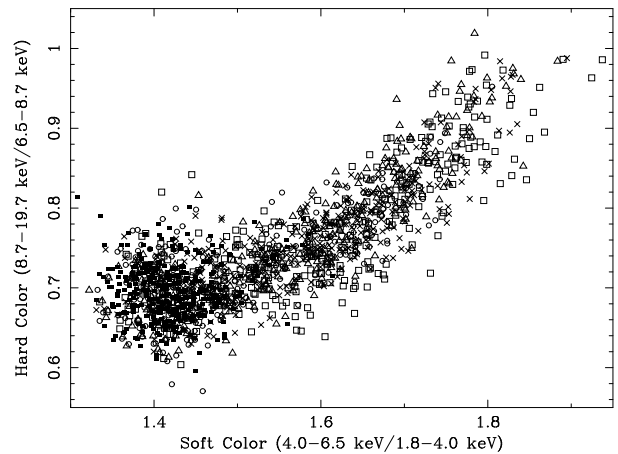}
\end{tabular}
\caption{The color-color diagram for the observation. 
Data points follow the same convention as in Fig.~4a.}
\end{center}
\end{figure}

We fine-tuned our measurement of the periodicity in the X-ray data by
fitting sine waves to the background-subtracted light curve in Fig.~1,
and determining the error on the
fitted parameters, producing a final period determination of
$P$=29377$\pm$40s =8.160$\pm$0.011 hrs. This is startlingly close to
the candidate orbital period suggested by Callanan et al.\ (1990) of
8.160$\pm$0.024 hrs, based on optical photometry.  While the close
agreement of our analysis with a ten-year-old optical result suggests
that the 8.16-hr periodicity is a stable and fundamental cycle of the
system, we clearly cannot rule out the possibility of a sequence of
flares intrinsic to the source with a fortuitous spacing.

The X-ray lightcurve (excluding the fifth observation) folded on this
period is shown in Fig.~3. The epoch of phase zero (defined for the
purposes of this paper as the ascending node of the sine wave) is
JD~245~0785.279$\pm$0.003. The modulation semi-amplitude is 24\% over
the entire 1.8--19.7 keV range used, increasing linearly with energy
from 14\% in the 1.8-4.0 keV range to 40\% at 8.7--19.7 keV. The
modulation during the fifth observation is greatly reduced but still
significant, at 3.8$\pm$0.1\% over the 1.8-19.7 keV band.
Considerable short-timescale ($\sim$minutes) variability is observed
in addition to the $\sim$8~hr modulation. Visual inspection of the
folded light curve (and detailed study of the individual light curves)
indicates that the source shows greater intrinsic X-ray variability at
phases $\phi$=0.0--0.5 than it does at later phases. To quantify this 
impression,
we subtracted the much longer-timescale 8.16-hr modulation
from the data and subjected the resulting light curve to a simple
statistical analysis: the weighted variance of the folded data is
$\sigma^2$=961.9 between $\phi$=0.0--0.5, as compared with
$\sigma^2$=129.4 between $\phi$=0.5--1.0.

In Fig.~4 we present the hardness-intensity diagram (HID) and
color-color diagram (CD) for the whole observation. The correlation
between hardness and intensity is very tight throughout the
observations. The CD shows a single straight, cohesive track with a
slight hook at the bottom left corner, perhaps indicating a vertex
with another branch. For the first four observations the individual
hardness-intensity and color-color diagrams are indistinguishable in
extent, scatter, and slope; for the fifth observation, the data points
cluster in the lower portions of the respective diagrams.

\begin{figure}[htb]
\figurenum{5}
\begin{center}
\begin{tabular}{c}
\psfig{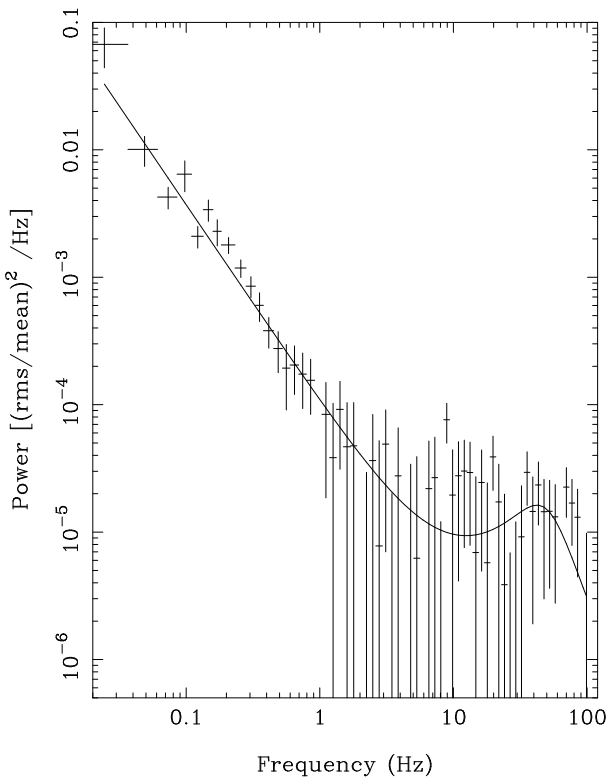}
\end{tabular}
\caption{The average power spectrum for \source\ in the energy range
2--10 keV. The Poisson level has been subtracted, and the data fitted
using a power law plus Lorentzian to characterize the VLFN and HFN
components.}
\end{center}
\end{figure}

\subsection{Fast timing characteristics}

The overall PSD of the source from 0.01--100~Hz can be well
represented using a single power law representing very low frequency
($\lesssim$1~Hz) noise (VLFN), with shape P $\propto$ $\nu^{-\gamma}$,
plus a higher-frequency noise component which we modeled using (i) a
cut-off power law, representing standard high frequency
($\gtrsim$10~Hz) noise (HFN),
or (ii) a Lorentzian, representing peaked HFN. First we examined the
mean power spectrum of all the observations, and then investigated it
as a function of position in the HID. Fits were performed in
several energy ranges: 2--6 keV, 6--20 keV, 20--100 keV, and 2--10
keV. The rms values are measured between 0.01--1 Hz for the VLFN and
0.01--100 Hz for HFN, and errors are $\Delta\chi^2$=1.0.

In the 2--6 keV power spectrum the VLFN component dominates between
0.01--1~Hz, with best fitting parameters of $\gamma$=1.50$\pm$0.07,
rms=4.5$\pm$0.3\%.  The best-fitting HFN power law has an index of
$\gamma_H$=0.09$^{+0.26}_{-0.18}$, rms=4.1$\pm$0.4\%. (The cut-off
frequency is found to be beyond the data range.)  Adding this second
power law improves the quality of the fit from $\chi^2/dof$=105/61
(VLFN only) to $\chi^2/dof$=81/59 (VLFN$+$HFN).  In the Lorentzian
representation, the center of the peaked noise was found to be
$\nu_H$=40$\pm$8~Hz, with FWHM=62$^{+43}_{-21}$~Hz, rms=5.1$\pm$0.9\%,
$\chi^2/dof$=75/58.  We show a power spectrum and best VLFN$+$HFN
model fit in Fig.~5.

At higher energies, VLFN continues to dominate. In the 6--20 keV
spectrum we find $\gamma$=1.56$\pm$0.11, rms=4.8$\pm$0.3\%,
$\chi^2/dof$=89/61, with upper limits of rms$<$1.5\% for simple
power-law HFN with $\gamma_H$ fixed at 0.0, and rms$<$1.4\% for peaked
HFN with $\nu_H$=40 Hz, FHWM=60 Hz. VLFN in the 20--100 keV range is
only slightly weaker, with $\gamma$=1.3$\pm$0.2, rms=3.9$\pm$0.3\%,
$\chi^2/dof$=77/61, with upper limits of rms$<$6.0\% on HFN.

We detect no quasi-periodic oscillations, and for typical $Z$-source
NB QPO, with centroid frequency $\sim$5~Hz and FWHM 2--3~Hz, we place
a 95\% confidence upper limit on the rms of 0.6\%. For parameters
typical of $Z$-source FB QPO (centroid frequency ranging from
$\sim$6~Hz to $\sim$20~Hz, with FWHM spanning $\sim$3--20~Hz) or atoll
source QPO (frequencies $\sim$1--60~Hz and similar FWHM) we find rms
upper limits of $\sim$1--3\%.

Power spectra accumulated for each of the first four observations
independently show no significant night-to-night change in the
strength or slope of the VLFN component, and insufficient statistics
to assess variability of the higher-frequency component.  We also
accumulated power spectra for the 2--10 keV data selected in several
broad color-intensity bands along the branch to search for behavioral
changes. At overall count rates $<$250 c/s we measure
$\gamma$=1.37$\pm$0.16, rms=3.2$\pm$0.9\%, increasing to
$\gamma$=1.63$\pm$0.15, rms=5.6$\pm$1.5\% at intermediate count rates,
and $\gamma$=1.49$\pm$0.10, rms=3.4$\pm$0.9\% at count rates $>$400
c/s.

\subsection{Spectral fitting}

The X-ray spectra from \source\ provide a means of parameterizing the
variations around the candidate binary cycle. We have examined the
summed 2--15 keV spectrum from the whole observation, and also the
results of dividing the data into ten spectral bins by phase.  The
summed spectrum can be well fit using a power law plus high energy 
cutoff model,
with photon index $\alpha$=0.37$\pm$0.04 and cutoff energy
$kT$=2.81$\pm$0.04 keV. In this model, the derived hydrogen column
density $N_H$ is indistinguishable from zero, with an upper limit of
10$^{21}$ cm$^{-2}$. With 1\% systematics added to represent the
current uncertainties in the response matrix generation, the goodness
of fit is $\chi^2/dof$=36.8/29. The summed data can also be fit using
a Comptonized Sunyaev \& Titarchuk model with
$N_H$=1.48$\pm$0.32$\times$10$^{21}$ cm$^{-2}$, electron temperature
$kT$=2.00$\pm$0.02 keV, and scattering depth $\tau$=14.2$\pm$0.4, with
a less satisfactory goodness-of-fit of $\chi^2/dof$=49.6/29. To a
certain extent these two models can be considered variations on a
theme, as both represent the scattering of cool photons on hot
electrons; simple Comptonization has been approximated with a cut off
power law $E^{-\alpha}$~exp($-E/kT$) in fits to LMXBs (e.g. White,
Peacock, \& Taylor 1985, White et al.\ 1986), while the Sunyaev \&
Titarchuk model uses the solution to the Kompaneets equation with
assumptions about the photon distribution through the scattering cloud
(Sunyaev \& Titarchuk 1980).

\begin{figure}[htb]
\figurenum{6a}
\begin{center}
\begin{tabular}{c}
\psfig{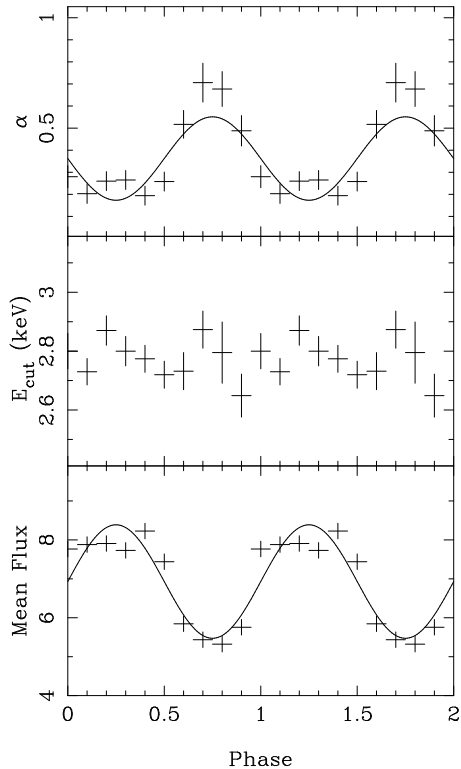}
\end{tabular}
\caption{Results of spectral fitting to a series of phase-resolved
spectra, showing the variation around the possible binary cycle of the
spectral fit parameters for (a) a power law plus high energy cutoff
model (b) a Comptonized Sunyaev \& Titarchuk model (see text).  The
mean flux per spectrum is given in units of
10$^{-10}$~erg~cm$^{-2}$~s$^{-1}$.  In each case two cycles are shown
for clarity.}
\end{center}
\end{figure}

\begin{figure}[htb]
\figurenum{6b}
\begin{center}
\begin{tabular}{c}
\psfig{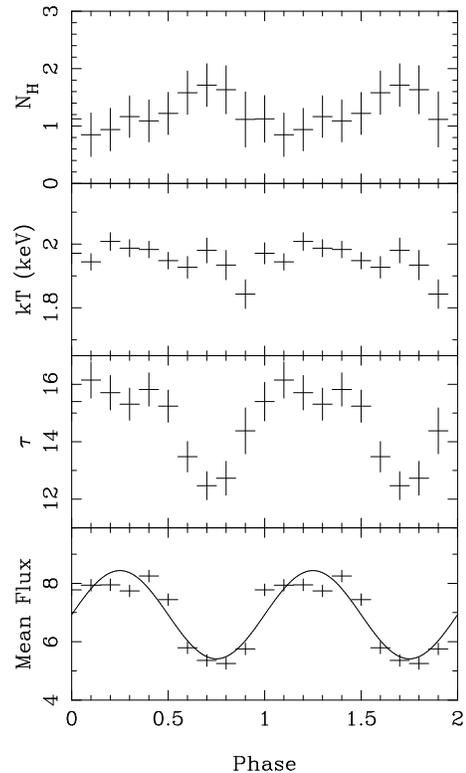}
\end{tabular}
\caption{See caption for Fig.~6a.}
\end{center}
\end{figure}

These two models were then fit to the ten phase-resolved
spectra from the source.  While the cut-off power law is a better fit
to the summed spectrum, this is not necessarily the case for the
individually-phased spectra; in fact, the Comptonized Sunyaev \&
Titarchuk model produced a better fit to the data for five out of the
ten cases. (Since the summed spectrum is effectively an average over
a series of different spectral states, the best-fitting model for
individual spectra may not be the model that best fits the summed spectrum.)
The variations of the fit parameters with phase are
illustrated in Fig.~6. In the first parameterization -- the power law
plus high energy cutoff -- the power
law index shows a clear correlation with phase, with the power law
slope being at its steepest at minimum light (as might be expected
from the larger modulation amplitude observed at higher energies). In
contrast, the value of the cutoff temperature does not vary
significantly with phase. In the second parameterization --
Comptonized Sunyaev \& Titarchuk -- the electron
temperature is similarly insensitive to phase, while the depth to
optical scattering $\tau$ shows a positive correlation. However, since
the upscattering is believed to take place in the immediate vicinity
of the neutron star, the physical interpretation of such a
dependence is not clear.

We have also fit a series of combination models to the data, chief
among them a cutoff power law plus blackbody, Bremsstrahlung plus
blackbody, a pair of blackbodies, and a disk-blackbody plus
blackbody. None of these combinations applied to the spectral data
provided an overall goodness-of-fit statistically superior to the
single-component Comptonization models, and in most cases the fits
themselves were insufficiently stable to determine meaningful error
estimates.  Of the combination models, the 
Bremsstrahlung plus blackbody fit provided the best
description of the data (despite being perhaps the least physically
realistic, due to the large emitting volumes required),
with parameters $kT_{Brems}$=4.5 keV, $kT_{BB}$=1.5 keV,
$\chi^2/dof$=32/28, with the blackbody making up $\sim$20\% of the
total flux. However, fitting the phase-selected spectra with this model we
found the fit parameters were not independent, in the sense that the
Bremsstrahlung temperature, the blackbody temperature, the derived
blackbody flux, and the proportion of the total flux made up by the
black body all increased with increasing flux in a correlated fashion.
This interdependency of unrelated fit parameters
is typical when fitting models that are
over-defined. We note that Bonnet-Bidaud et al.\ (1989) had similar
difficulties correlating the fit parameters for a Bremsstrahlung plus
black body model with the flux level of the source.

At the 50 kpc distance of the LMC, the mean observed flux translates
to a source luminosity of 1.8$\times$10$^{38}$ erg~s$^{-1}$,
in the middle of its known historical luminosity range.  For a
standard 1.4$M_\odot$\ neutron star binary with cosmic abundances 
this is equivalent to the Eddington luminosity.

\section{Discussion}

\subsection{\source: $Z$ or atoll?}

\source\ has never been explicitly assigned to either the $Z$ or the
atoll class of sources, but our results suggest that it is either a
bona fide $Z$-source, or an anomalous atoll source with many
$Z$-source characteristics.

The first consideration is the high luminosity of the source.  Atoll
sources have absolute luminosities that range from 0.01--0.1$L_{Edd}$
for the X-ray burst sources, up to 0.1--0.3$L_{Edd}$ for the bright GX
atoll sources (e.g. GX~9$+$9, GX~13$+$1).
The known $Z$-sources radiate consistently at a large
fraction of the Eddington luminosity, and frequently exceed that
luminosity for extended intervals. From their modeling of the spectral
variability of $Z$-sources, Psaltis et al.\ (1995) find
internally self-consistent solutions with accretion rates on the NB of
0.9--1.0$\dot M_{Edd}$, and rates $> \dot M_{Edd}$ on the FB.  The
luminosity of \source\ exceeds that of the bright atoll sources by a
factor of several, historically spanning the range
$\sim$0.4--2.0$L_{Edd}$, with a mean luminosity of $\sim$1.0$L_{Edd}$
at the time of our observations. \source\ also shows intervals of
flaring behavior (e.g. Bonnet-Bidaud et al.\ 1989; this paper), a
frequently-observed characteristic of many $Z$-sources.

Secondly, the CD and HID of \source\ show many similarities with
diagrams generated for confirmed $Z$-sources. In particular they
strongly resemble those sometimes observed from GX~349$+$2 in its
flaring branch (Hasinger \& van der Klis 1989; Kuulkers \& van der
Klis 1995), with a straight, tightly-correlated HID and an almost
equally straight and focussed CD. The concentration and distribution
of points in the lower left of the CD and HID in both sources suggest
the vertex with an incipient normal branch (seen explicitly in later
observations of GX~349$+$2: Kuulkers \& van der Klis 1998; Zhang,
Strohmayer, \& Swank 1998).  By contrast, most atoll sources show
pronounced curvature in their banana branch (Schulz et al.\ 1989,
Hasinger \& van der Klis 1989, Prins \& van der Klis 1997, among
others) and a more diffuse distribution of data points (see e.g. the
atoll plots constructed using RXTE PCA data from 4U1735$-$444,
Wijnands et al.\ 1998; 4U1608$-$52; M\'endez et al.\ 1999).  Among the
atoll-sources only the bright galactic center sources GX~9$+$9 and
GX~13$+$1 show a CD and HID comparable to that of \source\ (Hasinger
\& van der Klis 1989; Homan et al.\ 1998), and neither source does
this consistently (see e.g. Schulz et al.\ 1989).

$Z$ and atoll sources are classified using a combination of CD/HID
evidence and the results of fast timing analysis. The power spectra of
both source classes contain VLFN, with power law indices generally
$\gamma$=1.2--1.4 for atoll sources and $\gamma >$1.5 for $Z$-sources
in the flaring branch (e.g.\ Hasinger \& van der Klis 1989; Kuulkers
1995; Prins \& van der Klis 1997), although there are occasional
exceptions (e.g. KS1731$-$260, $\gamma$=1.5, Wijnands \& van der Klis
1997). The measured values for \source\ of $\gamma$=1.5--1.7 fall 
within the range observed for $Z$-sources. More compelling than
the absolute value of the power law index is the overall fast behavior
of the source as it travels along its branch. In the atoll sources the
power spectrum is generally dominated by HFN in the lower banana, and
by VLFN in the upper banana (e.g. van der Klis 1995; Prins \& van der
Klis 1997). By contrast, VLFN dominates all along the observed branch
of \source.

Close attention to the phenomenology of the VLFN is generally not a
high priority in studies of $Z$-sources, but we note that for Cyg~X--2
at high overall intensity, the LFN decreases and disappears on the
lower NB, while the VLFN remains strong and steep along the FB
(Wijnands et al.\ 1997). Steep power-law VLFN is similarly observed in
the low state of Cyg~X--2 (Kuulkers, Wijnands, \& van der Klis
1999). The VLFN in Sco~X--1 increases in strength as the source
travels from the NB/FB vertex to the end of the FB from rms=2\% to
6\%, with an almost-constant power-law index of
$\gamma_{Sco}$=1.72$\pm$0.01, while the HFN peaks at rms=3.0\% and
then fades into undetectability (Hertz et al.\ 1992). Along the FB,
Cir~X--1 is similarly dominated by VLFN (Shirey et al.\ 1998,
1999).  In both $Z$ and atoll sources there are indications that the
power law steepens as the source moves upward and rightward in the CD
(along the flaring branch, or from lower to upper banana), a trend
also (partially) observed in \source. Where observed, the HFN in
$Z$-sources typically has $\gamma\sim$0 and a cut-off frequency
approaching 100~Hz, consistent with the HFN observed in \source.  If
\source\ is a $Z$-source we might expect the presence of NBO/FBO; none
were found, although our upper limits are consistent with the
detection levels for such QPO in other $Z$-sources.

\subsection{The orbital periodicity}

We have detected an apparent strong 8.16-hr orbital periodicity in the
X-ray flux during four observations of \source, with an average
modulation amplitude of 24\% across the 1.8-19.7 keV energy band. In
the fifth observation, this amplitude has dropped to 4\% over the same
energy band.  This variability in the modulation depth, coupled with
the intrinsic variability of the source, may explain why the
periodicity is not evident in the public archival data from the
all-sky monitor instruments on RXTE, Vela 5-B, and Ariel V. The
pointed observations of \source\ from previous X-ray missions stored
in the HEASARC online archive are too short
to allow a search for a $\sim$8-hr periodicity;
the longest observation, performed using EXOSAT on 1985
December 10, lasts 18 hrs and shows 20\% variability, but no clear
orbital modulation.  The majority of LMXBs have orbital periods of
2--10 hrs. If the 8-hr period were confirmed as orbital, \source\
would fall towards the upper end of this distribution, implying a
dwarf companion of $\sim$1$M_{\sun}$ perhaps beginning to evolve off the
main sequence, with mass transfer starting to be driven by evolution
of the companion.

While the coincidence of the close agreement between the photometric
and X-ray periodicities is impressive, we strongly advise
caution. Given the sampling of our data we cannot rule out the
possibility of a fortuitous alignment of flare episodes, mimicking a
true period determination. We note that such an alignment previously
produced an erroneous period determination of 8.71 days in GX~349$+$2
(Ponman 1982). (The true period of 22.5 hr was identified from optical
photometry: Wachter 1997.)  

In addition, the physical mechanism that
might produce such a large modulation in the orbital light curve is by
no means clear.
At X-ray and optical energies, the shape of the observed light curves
in LMXBs, and the existence of an observable modulation, depend
strongly upon the inclination $i$, and the thickness and geometry of
the accretion disk.  At high inclinations ($i>80^\circ$) the central
source is obscured by the disk edge and the observed X-rays are
scattered through an accretion disk corona, producing a light curve
with a smooth sinusoidal modulation, interrupted by a partial eclipse
of the ADC by the companion (White \& Holt 1982, Mason \& C\'ordova
1982).  At inclinations (75$^\circ<i<$80$^\circ$) the central source
is often directly visible, and the $L_x/L_{opt}$ ratio thus
significantly enhanced. Sharp regular eclipses by the companion are
observed, along with deep irregularly-shaped dips caused by the
occultation of the central source by azimuthal structure on the
accretion disk associated with the impact point of the accretion
stream on the disk edge (e.g.\ X0748$-$676; Parmar et al.\ 1986).  At
70$^\circ<i<$75$^\circ$, X-ray dips are observed but no eclipses,
and optical light curves become roughly sinusoidal, with minima
occurring 0.2 in phase after the dips (e.g. X1254$-$690; Courvoisier
et al.\ 1986, Motch et al.\ 1987).  At inclinations lower than 70$^\circ$
we cease to see eclipses or dips, and the optical light curves are
generally sinusoidal (e.g. X1735$-$444, X1636$-$536, Corbet et al.\
1986, Smale \& Mukai 1988) with the exception of those sources in
which ellipsoidal variations are observed from giant companions.

The large-amplitude, quasi-sinusoidal X-ray modulation observed from
\source\ does not fit neatly into this picture. The deeper parts of
the folded light curve are not associated with a dramatic increase in
the $N_H$, as is generally the case for a source in which the
azimuthal structure plays a major role in defining the modulation, and
the amplitude of the modulation itself is clearly not stable,
indicating a volatile origin.  Neither sharp nor partial eclipses are
observed, thereby limiting the source inclination to $<$75$^\circ$. If
real, this would suggest that either the periodicity originates in:
(a) some modulation of the mass accretion rate itself (noting that the
variations in mass accretion are also responsible for the variations
of the source along the branches of the CD/HID), or (b) modulation of
the flux by the azimuthal disk structure according to the accepted
model for dipping sources, requiring that the source inclination falls
in the narrow 70$<i<$75$^\circ$ band. We note that the intrinsic X-ray
luminosity of \source\ is an order of magnitude greater than the
luminosities of the Galactic dipping sources, leaving us with no
precedent for such a system, and that no good physical model exists
for the height and distribution of the obscuring material at the
disk's edge even for the lower-luminosity dipping sources.

In previous work, evidence has been presented in favor of a 12.5-day
periodicity, based on optical photometry of \source\ (Crampton et al.\ 
1990). Radial velocity measurements are inconclusive, and this
longer period is not seen in the RXTE ASM light curve.  In LMXBs with
known periods in excess of a few days such as 4U0921$-$630 (8.99 days,
and Cygnus X--2 (9.84 days) the evolved giant companions can be
independently detected from the late-type absorption spectrum
underlying the strong emission-line spectrum from the accretion disk
(Cowley, Crampton, \& Hutchings 1982; Cowley, Crampton, \& Hutchings
1979), whereas spectroscopic studies of \source\ have as yet provided
no evidence for spectral features from the secondary star (Crampton et al.\
1990, Bonnet-Bidaud et al.\ 1989) even in the near-infrared, where we
might expect to see Ca~II absorption at $\lambda$8500\AA\ (Cowley 
et al.\ 1991).  However, we cannot at this time discount the possibility of
a 12.5-day modulation in the system, as superorbital periodicities
have been detected in several LMXBs (e.g. 175 days in 4U1820$-$303, 199
days in 4U1916$-$053, see Smale \& Lochner 1992, and references therein;
78 days in Cyg X--2, Wijnands, Kuulkers, \& Smale 1996) perhaps due to
precession of a tilted accretion disk. A program of long-term optical
photometry of \source\ is currently under way to resolve this question
(Wachter, Greene, \& Smale 1999).

An empirical relation exists between the absolute visual magnitude
$M_V$ of a LMXB, its mean X-ray luminosity as a function of
$L_{Edd}$, and its orbital period (van Paradijs \& McClintock
1994). (Such a relationship is expected from a simple model where the
bulk of the optical emission comes from -- and scales with the size of
-- a standard accretion disk.) Given a $V$-magnitude of 18.5 for
\source\ and a reddening of $E_{B-V}$=0.1 (which agrees well with the
$N_H$ expected from interstellar absorption in the direction of the
LMC; Bonnet-Bidaud et al.\ 1989), the derived absolute magnitude of
$-$0.3 leads to an expected orbital period of $\sim$17$^{+10}_{-8}$~hrs.
If our detected 8-hr periodicity does not stand the test of
time, future (X-ray or optical) search strategies should still be
focused on the 8--30 hr domain.

\section{Conclusion}

From this first study of the fast timing and CD/HID properties of
\source\ we have accumulated evidence that suggests it might be a
$Z$-source.  Considerations of the sustained near-Eddington and
super-Eddington luminosity, the shape and other characteristics of the
color-color diagram, and the fast time variability over the
0.01--100~Hz range all seem to support the hypothesis. If true, we
might expect to observe different branches of the $Z$ in future
observations of the source, perhaps while in intensity states
differing from those observed here.

$Z$-sources may fall into two classes, based on their inclination
(e.g.\ Kuulkers \& van der Klis 1995).  Sco~X--1, GX~349$+$2, and
GX~17$+$2 show stable $Z$s in their CDs and increase in brightness
along the FB, and may be observed at low inclination.  By contrast,
Cyg~X--2, GX~340$+$0, and GX~5$-$1 all show pronounced secular
variations in the positions of their $Z$s in the CD, and their
brightnesses show decreases or only slight increases along the FB
(where observed); these sources are believed to be observed at high
inclination.  If a $Z$-source, \source\ would probably join the first
class based on its FB behavior, adding even further to the difficulty
of creating a sustained high-amplitude orbital modulation in its X-ray
light curve.

\acknowledgments

This research has made use of data obtained through the High Energy
Astrophysics Science Archive Research Center Online Service, provided
by the NASA Goddard Space Flight Center. We acknowledge helpful
conversations with Rudy Wijnands, Stefanie Wachter, Phil Charles, \&
Katie McGowan.

\end{document}